\documentclass[aps,pre,reprint,longbibliography,superscriptaddress,amsmath,amssymb,twocolumn]{revtex4-2}
\usepackage[cp1251]{inputenc} 
\usepackage{amsmath}
\usepackage{amssymb}
\usepackage{bm}
\usepackage{datetime}
\usepackage[normalem]{ulem}
\usepackage{graphicx,color}
\usepackage{graphicx,scrtime}
\usepackage{tcolorbox}

\newcommand{\eps}{{\varepsilon}}

\newcommand{\ttau}{{\bm \tau}}

\newcommand{\kk}{{\bm k}}

\newcommand{\uu}{{\bm u}}

\newcommand{\rr}{{\bm r}}

\newcommand{\RR}{{\bm R}}

\newcommand{\LL}{{\cal L}}

\newcommand{\VV}{{\cal V}}

\newcommand{\tr}{\mbox{Tr}}
\newcommand{\be}{\begin{equation}}
\newcommand{\ee}{\end{equation}}
\newcommand{\ba}{\begin{eqnarray}}
\newcommand{\ea}{\end{eqnarray}}
\newcommand{\bse}{\begin{subequations}}
\newcommand{\ese}{\end{subequations}}
\newcommand{\beq}{\begin{eqnarray}}
\newcommand{\eeq}{\end{eqnarray}}
\newcommand{\ds}{\displaystyle}
\newcommand{\im}{\mbox{Im}}

\newcommand{\scs}{\scriptstyle}

\newcommand{\eeta}{\bm\eta}

\newcommand{\sk}{

\vspace{.2cm}

}

\newcommand{\new}[1]{\textcolor{black}{#1}}

\begin{document}
\title{Further testing the validity of generalized heterogeneous-elasticity theory for low-frequency excitations in structural glasses.}
\author{Walter Schirmacher}
\affiliation{Institut f\"ur Physik, Staudinger Weg 7, Universit\"at Mainz, D-55099 Mainz, Germany}
\affiliation{Center for Life Nano Science @Sapienza, Istituto Italiano di Tecnologia, 295 Viale Regina Elena, I-00161, Roma, Italy}
\author{Matteo Paoluzzi}
\affiliation{Dipartimento di Fisica, Universita' di Roma ``La Sapienza'', P'le Aldo Moro 5, I-00185, Roma, Italy}
\author{Felix Cosmin Mocanu}
\affiliation{Dept. of Materials, Univ. of Oxford, Parks Road, Oxford OX13PH UK}
\author{Dmytro Khomenko}
\affiliation{Dipartimento di Fisica, Universita' di Roma ``La Sapienza'', P'le Aldo Moro 5, I-00185, Roma, Italy}
\affiliation{NOMATEN Centre of Excellence, National Center for Nuclear Research, ul. A. Soltana 7, 05-400 Swierk/Otwock, Poland}\author{Gregorsz Szamel}
\affiliation{Dept. of Chemistry, Colorado State University, Fort Collins, CO 80523, USA}
\author{Francesco Zamponi}
\affiliation{Dipartimento di Fisica, Universita' di Roma ``La Sapienza'', P'le Aldo Moro 5, I-00185, Roma, Italy}
\author{Giancarlo Ruocco}
\affiliation{Center for Life Nano Science @Sapienza, Istituto Italiano di Tecnologia, 295 Viale Regina Elena, I-00161, Roma, Italy}
\affiliation{Dipartimento di Fisica, Universita' di Roma ``La Sapienza'', P'le Aldo Moro 5, I-00185, Roma, Italy}

\begin{abstract}
  \new{We summarize the salient features of our theory of non-phononic vibrational excitations in glasses 
    [W. Schirmacher {\it et al.}, Nature Comm. {\bf 15}, 3107 (2024)]. Next, we provide further evidence
    of the non-universality of the  $\omega^4$ scaling of the non-phononic vibrational density of states (DoS),
    and the existence of an important class of non-phononic
    excitations in glasses, which we call defect states. These modes are induced by frozen-in stresses and 
    can be classified as quasi-localized. Our results suggest that the commonly observed low-frequency $\omega^4$
    scaling of the non-phononic vibrational density of states is highly dependent on technical aspects of
    the molecular dynamics simulations employed to compute the DoS.} 
\end{abstract}
\maketitle
\new{
	\section{Introduction}}
\new{
The vibrational properties of glasses are until now subject to
extensive experimental, theoretical and simulational investigations
due to their anomalous features -- as compared to crystals -- and
the difficulty in treating a non-crystalline solid theoretically
\cite{elliott84,binder05}. 
The low-frequency vibrational properties of glasses and their
associated low-temperature thermal properties differ appreciably
from those of their crystalline counterparts
\cite{ramos22}.
Experimental information relies
on inelastic light scattering (Raman and Brillouin, \cite{berne76}), inelastic neutron
\cite{schober14} and
X-ray scattering \cite{baron20}, and indirectly via the specific
heat and thermal conductivity \cite{ramos22}.
The most striking difference between the vibrational density of states
(DoS) $g(\omega)$ and that of a crystal is an enhancement with respect to
Debye's $g(\omega)\propto \omega^2$ law (``boson peak''\footnote{See \cite{ramos22,schirmacher24b} for an extensive discussion of the boson peak.}). This anomalous feature appears
as a shoulder in the DoS and a peak in $g(\omega)/\omega^2$.}

\new{Macroscopically glasses, and, more generally, disordered solids support
acoustic waves like crystals, which therefore are generic low-frequency
excitations, which phenomenologically may be described in both crystals
and glasses by elasticity theory. It has been demonstrated in a series
of scientific works \cite{schirm06,schirm07,marruzzo13,kohler13,schirm14} 
that {\it spatially fluctuating elastic constants} give rise to a deviation
from Debye's $\omega^2$ law due to the phonons. This has been verified by
calculations based on the self-consistent Born approximation (SCBA) \cite{schirm06},
the coherent-potential approximation (CPA) \cite{kohler13}, as well by
numerical simulations \cite{marruzzo13}. This approach is known as heterogeneous-elasticity theory (HET).}

\new{Subsequently, several groups published results of numerical simulations of
systems, which are so small that they are not able to support elastic
waves \cite{Lerner2016,Lerner2017,parisi18,parisi19,lernerbouchbinder21,parisi21}.
A class of anomalous modes was identified, whose DoS scales as a power-law of the frequency.
However, some disagreement was found concerning the precise value of the scaling exponent.
More specifically, in a
representation $g(\omega)\propto\omega^s$ one group of authors
obtained universally $s=4$, independently on the sample preparation
\cite{Lerner2016,Lerner2017,lernerbouchbinder21}, whereas the authors of
\cite{parisi19,parisi21} found $2\leq s \leq 4$, depending on the
parental temperature $T^*$ in the liquid state, from which the
glassy sample was obtained by quenching. For high $T^*$ they found
$s\approx 2$, for low $T^*$ $s\approx 4$.} 

\new{In order to make further progress, we developed a generalized
elasticity theory (GHET), in which, in addition to spatial fluctuations of
elastic constants, spatially fluctuating frozen-in stresses were taken
into account \cite{schirmacher24}. These stresses were found to couple to vortex-like displacement
fields. These non-phononic vibrational excitations were termed {\it type-II},
whereas those due to spatially fluctuating elastic constants
{\it type-I}. With the help of GHET the parental-temperature scenario was 
explained in the following way: For rather high $T^*$ the system is
in a state of marginal stability, for which HET predicts a
DoS with $s=2$, which is not a Debye exponent but a critical one.
For stable systems HET (without the type-II modes)
predicts \mbox{-- for small samples without phonons --}
a low-frequency gap in the spectrum. Within GHET this gap is filled with the 
type-II excitations. The DoS of these type-II excitations depends on the distribution
of the internal stresses. This distribution of internal stresses, and hence the DoS, was found to be
 sensitive to the way the potential is smoothed around its cutoff (tapering procedure).}

Recently, in a letter to the Editor of Phys.~Rev.~E~\cite{lerner25}, E. Lerner and E. Bouchbinder \new{raised} concerns about the results \new{of} Ref.~\cite{schirmacher24}. Specifically, they insist on the universality of  a $g(\omega)\propto \omega^4$ scaling of the low-frequency non-phononic density of states for glasses, and suggest that our theoretical treatment would be ``inherently deficient in capturing the .. nature of quasilocalized, nonphononic excitations in structural glasses''.

\new{In this paper we first
present briefly the derivation and the main results of GHET,
in particular focussing on the role of frozen-in stresses.
We then present further evidence of our findings and discuss them in the context of concerns raised by Lerner and Bouchbinder.}
\sk\sk
\new{
\section{Generalized heterogeneous-elasticity theory}
\subsection{Derivation}
In our derivation of GHET \cite{schirmacher24}, we started from the expression for the harmonic
potential energy of a system interacting via a pairwise
potential $\phi(r)$
\be\label{e1}
E_{\rm harm}\{\rr_i\}
=\sum_{(i,j)}
\uu_i {\cdot}  \stackrel{\leftrightarrow}{H}_{ij} {\cdot} \uu_j
\doteq
\sum_{(i,j)}V_{ij}
\ee
Here $\sum\limits_{\scs(i,j)}$ denotes a sum over pairs $(i,j)$, $\uu_i$ are
infinitesimal displacements from the disordered locations $\rr_i$
of the atoms in the glass, and $\stackrel{\leftrightarrow}{H}_{ij}$
is the dynamical (Hessian) matrix given in terms of force
constants $K_{ij}^{\alpha\beta}$
\ba\label{hessian1}
H_{ij}^{\alpha\beta}&=&
\frac{\partial}{\partial r_i^\alpha}
\frac{\partial}{\partial r_j^\beta}
E_{\rm harm}\{\rr_i\}
\nonumber\\
&=&-K_{ij}^{\alpha\beta}\big(1-\delta_{ij}\big)
+\bigg(\sum_{\ell\neq i}
K_{i\ell}^{\alpha\beta}\bigg)\delta_{ij}
\ea
The force constants are given by
\ba\label{forcec}
K_{ij}^{\alpha\beta}&=&\bigg[\phi''(r_{ij})-\frac{1}{r_{ij}}\phi'(r_{ij})\bigg]
\frac{
        r_{ij}^\alpha r_{ij}^\beta}{r_{ij}^2}
+\frac{1}{r_{ij}}\phi'(r_{ij})\delta_{\alpha\beta}\nonumber\\
&\doteq&{\phi^{(1)}\!(r_{ij}) }\;
r_{ij}^\alpha r_{ij}^\beta \;  + \; {\phi^{(2)}\!(r_{ij}) }\; \delta_{\alpha\beta}
\ea
with implicit definition of the functions $\phi^{(1)}\!(r_{ij}) $ and $\phi^{(2)}\!(r_{ij})$.}

\new{In order to derive the disordered version of elasticity theory
from the microscopic Hamiltonian, we cannot use the standard
procedure of lattice dynamics \cite{ashcroft76a}, which is based on
the translational symmetry of crystals. Instead we follow
the ideas of Lutsko \cite{lutsko88,lutsko89} and Alexander \cite{alexander98}, using the fact
that $E_{\rm harm}$ only depends on the distances between
the atoms. We therefore 
introduce difference and
center-of-mass variables
$\rr_{ij}=\rr_i-\rr_j$.
\mbox{$\RR_{ij}=\frac{1}{2}\big[\rr_i+\rr_j\big]$.}
We interpret $\RR_{ij}\doteq\RR$ as the local vector of the
continuum theory and define the potential-energy
density $\VV$ as
\ba
\VV(\RR)&=&\frac{V_{ij}}{\Omega_Z}\nonumber\\
&=&
\frac{1}{2\Omega_Z}\bigg(
{\phi^{(1)}\!(r_{ij}})
\sum_{\alpha\beta}
r_{ij}^\alpha
r_{ij}^\beta
u_{ij}^\alpha
u_{ij}^\beta
+ {\phi^{(2)}\!(r_{ij})} \;
u_{ij}^2\bigg)\nonumber\\
&\doteq&\VV^{(1)}(\RR)
+\VV^{(2)}(\RR)\, ,
\ea
where
$\uu_{ij}=\uu_i-\uu_j$ and $\Omega_Z$ is the inverse
of the average number of bonds $(i,j)$ per volume, given by
$\frac{1}{\Omega_Z}=\frac{1}{2\Omega}N(Z-1)$. Here $\Omega$
is the total volume, $N$ the number of atoms, and
$Z$ the average coordination number.
It is important to note that $\sqrt[3]{\Omega_Z}$ is the minimal
possible length scale to formulate a continuum theory.
We make a Taylor expansion of $\uu(\rr_i)$ around $\RR$
\ba\label{ui}
\uu(\rr_i)&=&\uu(\RR)+\big([\rr_i-\RR]\cdot\nabla\big)\uu(\RR)\nonumber\\
&=&\uu(\RR)+\frac{1}{2}\big(\rr_{ij}\cdot\nabla\big)\uu(\RR)
\ea
It follows
\be
\frac{1}{2}\big[\uu {(\rr_i)}+\uu{(\rr_j)} \big]=\uu(\RR)
\ee
and
\be\label{uij}
u^\alpha(\rr_i)-
u^\alpha(\rr_j) =
\sum_\gamma r_{ij}^\gamma
u_{\alpha|\gamma}(\RR)=u_{ij}^\alpha\, ,
\ee
with  abbreviation
$u_{\alpha|\gamma} {\doteq}  \;  \partial_\gamma u^\alpha$.
The two terms of the potential-energy density become then
\ba
\VV^{(1)}(\RR)
&\doteq&\frac{1}{\Omega_Z}V_{ij}^{(1)}
        \bigg|_{\RR=\RR_{ij}}\\
&=&\frac{1}{2}\sum_{\alpha\beta\gamma\delta}
B^{\alpha\beta\gamma\delta}(\RR)
\eps^{\alpha\gamma}(\RR)
\eps^{\beta\delta}(\RR)\nonumber\\
\VV^{(2)}(\RR)
&\doteq&\frac{1}{\Omega_Z}V_{ij}^{(2)}(\RR)
        \bigg|_{\RR=\RR_{ij}}\\
&=&
\frac{1}{2}\sum_{\alpha\gamma\delta}
\sigma^{\gamma\delta}(\RR)
u_{\alpha|\gamma}(\RR)u_{\alpha|\delta}(\RR)\nonumber
\ea
with the strain tensor
\be\label{strain}
\eps^{\alpha\gamma}(\RR)
=\frac{1}{2}\big[
        u_{\gamma|\alpha}(\RR)
+u_{\alpha|\gamma}(\RR)
        \big]\, ,
\ee
the Born-Cauchy elastic constants 
\be
B^{\alpha\beta\gamma\delta}(\RR)
{\doteq}
\frac{1}{\Omega_Z}
{\phi^{(1)}\!(r_{ij})  \; }
r_{ij}^\alpha
r_{ij}^\beta
r_{ij}^\gamma
r_{ij}^\delta\bigg|_{\RR=\RR_{ij}}\, ,
\ee
and the local stresses
\be
\sigma^{\gamma\delta}(\RR)
{\doteq}
\frac{1}{\Omega_Z}\
{\phi^{(2)}\!(r_{ij}) \; }
r_{ij}^\gamma
r_{ij}^\delta
\bigg|_{\RR=\RR_{ij}}\, .
\ee
}
\new{Here, $\VV^{(1)}$ 
is the stress-independent local potential energy density
with spatially fluctuating elastic constants, whereas $\VV^{(2)}$ is the
term due to the frozen-in stresses $\sigma^{\alpha\beta}(\RR)$.}

\new{It has been noted by Lutsko \cite{lutsko88,lutsko89}, that the
local stresses affect the local elasticity. More importantly,
Alexander \cite{alexander84,alexander98} noted that the stress-related
terms {\it violate local rotational invariance}. Therefore the stress-related terms not only involve the {\it strain tensor} in Eq.~(\ref{strain}), but
also the {\it rotation tensor}
\be\label{rotation}
\eta^{\alpha\gamma}(\RR)
=\frac{1}{2}\big[
        u_{\gamma|\alpha}(\RR)
-u_{\alpha|\gamma}(\RR)
        \big]\, .
\ee
The tensor $\eta^{\alpha\gamma}$ has only three entries, and
is related to the {\it vorticity vector} $\eeta$ by
\be\label{vorticity}
\eta_\alpha(\RR)=
        \sum_{\beta\gamma\delta}\lambda_{\alpha\beta\gamma}
        \eta_{\beta\gamma}(\RR)\, ,
\ee
where $\lambda_{\alpha\beta\gamma}$ is the Levi-Civita symbol.
Following \cite{lutsko88}, we incorporate the symmetrized
stress terms into the elastic constant tensor as
\be
C_{ij}^{\alpha\beta\gamma\delta}=
B_{ij}^{\alpha\beta\gamma\delta}+
{\frac{1}{4}    \big(
\sigma^{\gamma\delta}\!\delta_{\alpha\beta}+
\sigma^{\alpha\delta}\!\delta_{\gamma\beta}+
\sigma^{\gamma\beta}\!\delta_{\alpha\delta}+
\sigma^{\alpha\beta}\!\delta_{\gamma\delta}
        \big)}
	\ee
This modifies the two terms of the local potential-energy density
$\VV(\RR)\rightarrow
\widetilde\VV^{(1)}(\RR)+
\widetilde\VV^{(2)}(\RR)$
with
\be
\widetilde\VV^{(1)}(\RR)=
\frac{1}{2}C^{\alpha\beta\gamma\delta}(\RR)
\eps^{\alpha\gamma}(\RR)
\eps^{\beta\delta}(\RR)
\ee
and
\be\label{e17}
\widetilde\VV^{(2)}(\RR)=
\frac{1}{2}\sum_{\alpha\gamma\delta}
\sigma^{\gamma\delta}(\RR)\eta^{\alpha\gamma}(\RR)\big(
\eta^{\alpha\delta}(\RR)-
\eps^{\alpha\delta}(\RR)\big)
\ee
In terms of the vorticity vector $\eeta$, Eq. (\ref{e17}) can
be written as
\ba
\widetilde\VV^{(2)}(\RR)
&=&\frac{1}{2}\bigg(\tr\{\sigma(\RR)\}\eeta^2-
	\sum_{\gamma\mu}\sigma^{\gamma\delta}(\RR)
	\eta^\gamma(\RR)
	\eta^\delta(\RR)\bigg)\nonumber\\
&&\qquad-\frac{1}{2}\ttau\cdot\eeta
\ea
with the coupling vector
\be\label{coupling}
\tau_\alpha=\sum_{\beta\gamma\delta}\lambda_{\alpha\beta\gamma}
\sigma^{\beta\delta}
\eps^{\delta\gamma}
\ee
\subsection{Type-I spectrum: boson peak and marginal stability}
Neglecting the $\VV^{(2)}$ term, we have obtained a microscopic
derivation of heterogeneous-elasticity theory
\cite{schirm06}, which, as mentioned above, predicts a boson peak
as a consequence of disorder. The associated vibrational modes 
feature a spectrum according to the Gaussian Orthogonal random-matrix
ensemble (GOE) \cite{mehta67}. They
are random-matrix modes (eigenfunction of random matrices), and
have been called ``type-I'' nonphononic modes in Ref. \cite{schirmacher24}.
In this paper the type-I modes have been shown numerically 
to
feature a level-distance
spectrum according to the Gaussian Orthogonal random-matrix
ensemble (GOE) \cite{mehta67}, and are therefore delocalized. 
Only at very high frequencies the states are localized.
It is known from earlier work,
that at very high frequencies, near the Debye frequency, the eigenstates
of a disordered harmonic system are localized \cite{schirm98,allen99}.}

\new{In the SCBA version of HET the underlying Gaussian distribution of
elastic constants leads to an instability at a critical value
of the disorder, characterized by the relative variance of the
fluctuations. At marginal criticality the boson peak extends to
zero frequency, and the DoS is quadratic in $\omega$ like in the
case of the Debye phonons, albeit with a higher prefactor. 
In the discussion of non-phononic modes of very small systems,
which do not allow for standing waves, we have argued
\cite{schirmacher24} that the observed $g(\omega)\propto \omega^2$
behavior of samples quenched from very high parental temperatures
is due to marginal stability. 
In more stable system SCBA-HET predicts a gap below the boson
peak. This gap is then filled, in the presence of local frozen-in
stresses, by ``type-II'' modes, which are governed by
the second term $\widetilde \VV^{(2)}$of the potential energy density.}

\new{
	\subsection{Type-II spectrum: The role of frozen-in stresses}
As in Ref. \cite{schirmacher24} we specify our model now in
assuming that the local stresses are confined to a certain
volume $\Omega_\ell$ with center at $\RR_\ell$, and we give
the label $\ell$ to these stresses and the corresponding
vorticities. 
}

\new{The total Lagrangian density of GHET is 
given by
\be
\LL_{\rm GHET}(\RR,t)=\LL_{\rm HET}(\RR,t)+\frac{1}{2}\zeta\sum_\ell\dot\eeta_\ell^2(\RR,t)-\VV^{(2)}(\RR,t)\, .
\ee
where $\zeta$ is an average moment-of-inertia density, and the
HET Lagrangian is
\be
\LL_{\rm HET}(\RR,t)=
\frac{\rho}{2}[\dot u(\RR,t)]^2
-\widetilde\VV^{(1)}(\RR,t)\, .
\ee
We treat now $\uu(\RR,t)$ and $\eeta(\RR,t)$ as independent variables
and obtain the following Lagrangian equations of motion in frequency space
($z=\omega^2+i\epsilon$)
\ba\label{eqmo1}
-\zeta z \; \eta_\ell^\nu(\RR,z)&=&
-\tr\{\sigma\}\eta_\ell^\nu(\RR,z)
+\sum_{\mu} \sigma_\ell^{\nu\mu}(\RR)
\eta_\ell^{\mu}(\RR,z)\nonumber\\
&&\qquad 
-\tau^\nu_{\ell}(\RR,z)\, ,
\ea
\ba\label{eqmo2}
-\rho_m z\; u^\alpha(\RR,z)
&=&\sum_\beta\frac{\partial}{\partial x_\beta}\bigg(\sum_{\gamma\delta}C^{\alpha\gamma\delta\beta}(\RR)\eps^{\gamma\delta}(\RR,z)\nonumber\\
&&+\sum_{\nu}\sum_{\ell}s^{(\nu),\alpha\beta}_{\ell}(\RR)\eta^\nu_\ell(\RR,z)\bigg)\, ,
\ea
with
\be\label{coupling1}
\tau^\mu_\ell(\RR,z)=\sum_{\beta\gamma\delta}\lambda_{\mu\beta\gamma}
\sigma^{\beta\delta}(\RR)
\eps^{\delta\gamma}(\RR,z)\, ,
\ee
\be
s^{(\nu),\alpha\beta}_{\ell}(\RR)=\frac{1}{2}\left\{
        \begin{array}{cc}
		t^{(\nu),\alpha\alpha}_{\ell}(\RR)&\alpha=\beta\\
		{\ds\frac{1}{2}}t^{(\nu),\alpha\beta}_{\ell}(\RR)&\alpha\neq\beta
        \end{array}\right.
	\ee
and
\be
t^{(\nu),\alpha\beta}_{\ell}(\RR)=\frac{\partial}{\partial \eps^{\alpha\beta}}
\ttau^{\nu}_{\ell}
=\sum_\gamma\lambda_{\nu\gamma\alpha}\sigma^{\beta\gamma}_\ell(\RR)
\ee
Eqs. (\ref{eqmo1}) and (\ref{eqmo2}) comprise the GHET equations
of motion.\sk
In order to gain qualitatively insight into the influence of the
stresses on the type-II non-phononic spectrum, we make the following
simplifications
\begin{itemize}
	\item[-] We disregard the $\RR$ dependence of the elastic constants,
		i.e. disregard the type-I nonphononic vibrations;
	\item[-] We treat the $\eeta$ and $\sigma^{\alpha\beta}$
		as scalars $\eta_\ell$ and $\sigma_\ell$~~\footnote{The stresses $\sigma_\ell$ may take both signs as a consequence of the external
		pressure imposed by the boundary conditions, see \cite{schirmacher24}.};
	\item[-] We assume that inside of the vorticity range
		$\Omega_\ell$, the stresses $\sigma_\ell$ do not depend on $\RR$;
	\item[-] We assume that the type-II modes couple only to the
		transverse elastic waves $\uu_T(\kk,z)$, which are supposed to
		propagate in direction of the $z$ coordinate, and polarized
		in $x$ direction.
\end{itemize}
		Eqs. (\ref{eqmo1}) and (\ref{eqmo2}), transformed
		into $\kk$ space, then simplify as follows:
\be\label{eqmo1a}
\bigg(-\zeta z + \sigma_\ell\bigg)\eta_\ell(\kk,z)=-\frac{1}{2}ik\sigma_\ell u_T(\kk,z)
\ee
\be\label{eqmo2a}
\bigg(-\rho z + k^2\mu\bigg)u_T(\kk,z)=ik\sum_\ell\frac{1}{4}\sigma_\ell\eta_\ell(\kk,z)
\ee
Here $\mu$ is the shear modulus.
Solving Eq. (\ref{eqmo1a})  for $\eta_\ell(\kk,z)$ and inserting
this into Eq. (\ref{eqmo1a}) we get
\be\label{eqmo3}
\bigg(
-\rho z + k^2\big[\mu-\Sigma(z)\big]
\bigg)u_T(\kk,z)=0
\ee
with
\be
\Sigma(z)=\frac{1}{8}\sum_\ell\frac{\sigma_\ell^2}{-\zeta z+\sigma_\ell}
\ee
This self-energy gives a frequency-dependent contribution to the
shear modulus. The modified
density of states of the transverse modes is given by
\be
g(\omega)=
\overline{
\frac{2\omega}{\pi}
\im\left\{
\sum_\kk
\frac{1}{
	-\rho z + k^2\big[\mu-\Sigma(z)\big]}
\right\}
}\, ,
\ee
where the overline denotes an average over the stresses
$\sigma_\ell$ with distribution density $P(\sigma)$. 
In addition to the contribution of the transverse waves,
which is absent in small enough samples, there is
a contribution
to the DoS from the frozen-in stresses
\be\label{ghetdos}
\Delta g(\omega)\propto \omega \im\{\Sigma(z)\}
\propto \omega \sigma^2\,P(\sigma)\bigg|_{\sigma=\zeta\omega^2}
\ee
}

\new{Eq. \eqref{ghetdos} implies that according to GHET the low-frequency part of the DoS reflects the distribution of small frozen-in stresses
in the glass. It follows that} the
frequency-dependence of the DoS, 
determined from molecular dynamics (MD) simulations, may depend on technical details of the simulational procedure. We have shown \new{in \cite{schirmacher24}
by our numerical simulations,}
that the low-frequency DoS -- in stable systems far from marginality, \new{in fact,} depends on the distribution of small stress values. In MD simulations, however, many low stress values are introduced by the tapering 
(smoothing) function. This function is used to ensure smooth vanishing of the potential at the cutoff distance. This cutoff is in turn introduced to reduce the number of interacting particles and thus the computational cost. 
The standard choice is a polynomial tapering function
which ensures that the
second derivative
of the potential is continuous at the cutoff.  According to the theory, this generates a stress distribution density $P(\sigma)$ near the cutoff that scales as $\sigma^{-\frac{1}{2}}$,  and this in turn produces a DoS scaling as $\omega^4$. 
However, the theory predicts that other choices of tapering function gives rise to an $\omega^s$ scaling with an exponent $s\neq 4$.

The numerical simulation reported in \cite{schirmacher24}, despite the usual limitation on the range of accessible frequencies, are consistent with this theoretical prediction.

\new{\section{Additional evidence of GHET predictions}}

In their letter \cite{lerner25}, Lerner and Bouchbinder raise five distinct concerns about our study \cite{schirmacher24}. In the following we will comment on these concerns and bring further evidence of the results of~\cite{schirmacher24}. We start with a brief summary of the five main results of Ref.~\cite{schirmacher24} on which concerns have been raised in~\cite{lerner25}:

	(i)
		In earlier simulations \cite{parisi19,parisi21} 
		it was shown that the DoS in systems quenched from
		high parental temperatures scales with an exponent $s$ = 2, while
		higher values of $s$ up to $s$ = 4 are observed when the system is quenched from lower parental temperatures.
		We have explained this finding by the observation 
		that in these studies the samples with high
		parental temperature were marginally stable, whereas
		the low-parental-$T$ samples were more stable.

	(ii)
In all our simulations the spectral statistics of the non-phononic
		eigenvalues obey the GOE (Gaussian-Orthogonal-Ensemble)
		statistics, from which follows that both the
		type-I and the type-II modes are delocalized.

	(iii)
In our simulations, for samples quenched from low parental temperatures, we find a sensitive dependence of the low-frequency DoS
 on tapering, in agreement with our theoretical prediction.

	(iv)
In our theory we predict that
in systems whose potential displays a minimum  (such as Lennard-Jones systems), an
exponent $s$ = 5 of the DoS scaling is predicted, which, however,
is modified by the tapering.

	(v)
We deduce from our analytic work that the type-II eigenfunctions
feature vortex-like patterns, which we observe in the simulations.

We now comment point by point on the concerns expressed in
Ref. \cite{lerner25} on these issues:

(i)
The authors of \cite{lerner25} do not find the \(\omega^2\) (\(s=2\)) law 
in their simulations of samples quenched from high parental temperatures,
which is expected for a marginally stable system \cite{franz15}.
On the other hand, in all our
simulations in Refs. \cite{parisi19,parisi21,schirmacher24} we have
clear evidence for $s=2$ in the case of quenching from high enough
parental temperatures.
In order to show the absence of $s=2$, and the universality of $s=4$, 
the authors of \cite{lerner25} report the integrated DoS
$F(\omega)=\int_0^\omega g(\tilde \omega)d \tilde \omega$, divided by
$\omega^{s+1}$ with $s=4$, see~\cite[Fig.~1]{lerner25}.

First, we note that the data in \cite[Fig.~1]{lerner25} never become really flat except, perhaps, in the small frequency region between 0.1 and 0.3 - which is a rather small range to fit a power-law with reasonable precision.
From the small slope observed in \cite[Fig.~1]{lerner25}, one can deduce that the exponent $s$ ranges from \(s=3\) (small systems) to \(s=3.5\) (larger systems). Furthermore, the data for $N=32768$ and $N=131072$ seem to be 
very close, indicating that one has reached convergence.
We can then conclude that the simulations of Ref.~\cite{lerner25}, consistently with other studies (see e.g.~\cite{wang21}),
find neither the claimed \(s=4\) nor the marginally stable value \(s=2\).

Second, to obtain a marginally stable system, one needs to (1)~start from a very high temperature, and (2)~quench it with algorithms that do not allow the system to relax towards ``comfortable" (i.e., far from marginality) situations. We cannot comment on the algorithm used to quench in \cite{lerner25}, nor on the fact that their parental temperature (``roughly four times larger than the glass transition temperature") is high enough, as these aspects are strongly system-dependent. A detailed study of the procedures to create marginal stable glasses, and of their DoS is underway.

We conclude that at present the origin of the discrepancy between our simulations, where $s=2$ is found at high parental temperatures, and those reported in~\cite{lerner25} remains unclear.

(ii)
In their simulations of small glasses, reported in~\cite{lerner25},
the authors evaluated
the participation ratio
\begin{equation}
e=\frac{1}{N}\big[
	\sum_i(
	{\bm\psi}_i
	\cdot
	{\bm\psi}_i
	)^2
	\big]^{-1}
\end{equation}
(where $N$ is the number of particles and ${\bm \psi}_i$ is an eigenvector
to an eigenvalue $\omega_i^2$) for several system sizes $N$
from $N$ = 2048 to $N$ = 131072. In Ref.~\cite[Fig.1b]{lerner25} the
quantity $Ne$ is reported, showing a dense cloud at low frequencies
with values ranging from
20 to 200. As this cloud of data is rather diffuse, showing no trend,
the authors conclude that $Ne$ would be constant, consequently
$e$ would scale as $1/N$, from which would follow that the corresponding modes
are localized.
In the footnote [55] the authors call the corresponding
modes quasilocalized.

In our study
\cite{schirmacher24} we evaluated the {\it spectral statistics} of the
eigenvalues both for systems quenched from high and from low parental temperatures.
The statistics obeys the GOE (Gaussian orthogonal ensemble) in all cases, 
which means that they exhibit level repulsion and are therefore
delocalized. 
This apparent contradiction can be solved by noting that vibrational modes in disordered systems can have very non trivial structure,
such that they can be delocalized in a subtle way~\cite{franz25}.

We would like to comment more generally on quasilocalized modes
in glasses. This term was coined by Schober and Oligschleger \cite{schober96}
who pointed out that local vibrational defect states are inevitably coupled
to the elastic degrees of freedom, leading eventually to hybridization with
phonons and to delocalization. This point was further investigated by
Schober and one of the present authors, in a systematic study of the
localization properties of
low-frequency vibrational states of small glassy systems 
upon varying system size \cite{schober04}. For all system
sizes studied ($N$ = 2048 to 32000) the participation ratio was reduced
at low frequencies ($e =0.4$) but did not depend on $N$. The eigenvalue
nearest-neighbour statistic, however, showed GOE behaviour, even for
$N = 2024$. As these findings are in agreement to those in
our recent investigation \cite{schirmacher24}, we conclude that,
in fact, what we call ``type-II'' can be also classified as
quasilocalized modes and are delocalized. It has, however,
been found in Ref. \cite{schober04} that the energy associated
with a quasilocalized mode is concentrated in a certain region
in space. This is what we also assume to be the case for the type-II modes
\cite{schirmacher24}.

(iii)
The authors of Ref. \cite{lerner25} do not agree with our conclusion that
the smoothing of the potential near its cutoff
$\phi(r)\propto (r_c-r)^{m+1}$ (tapering) strongly influences the DoS of a stable simulated small glass
at low frequencies.
To show this, they report in Ref.~\cite[Fig.~2c]{lerner25} {\it our} $F(\omega)$ data ($m$=2 and $m$=$\infty$), taken from~\cite{schirmacher24}.
Contrary to our finding ($s\approx4$ for $m$=2 and $s\approx3$ for $m$=$\infty$) they claim that the data are consistent with $s$=4 independently of $m$.
We note that this finding is based on an extremely small range of frequencies, $1.2 \le \omega \le 1.4$. 
In order to show that our $s$ values, and the corresponding conclusions
concerning the tapering, are correct, we collected new data for
$m$ = 2 and $m=\infty$.
We plot in Fig.~1 the quantities $F(\omega)/\omega^{5}$, panel (a) and $F(\omega)/\omega^{4}$, panel (b) for the two cases $m$=2 and $m$=$\infty$. 
We display
the first $5 \cdot 10^4$ eigenfrequencies of a set of order $10^6$ eigenmodes.
The figure shows that the low-frequency data are closer to a slope $s=4$ for $m=2$ and $s=5$ for $m=\infty$,
even if large uncertainties in the determination of $s$ are present. In any case, the DoS are clearly different at low frequencies.
We want to emphasize that plotting $F(\omega)$ and passing a straight line on the low-frequency data leads to a much larger uncertainty on $s$; testing whether the data agree with a given value of $s$ requires, in our opinion, plotting $F(\omega)/\omega^{s+1}$ as we did in Fig.~1. 

\begin{figure}[t]
\includegraphics[width=.95\columnwidth]{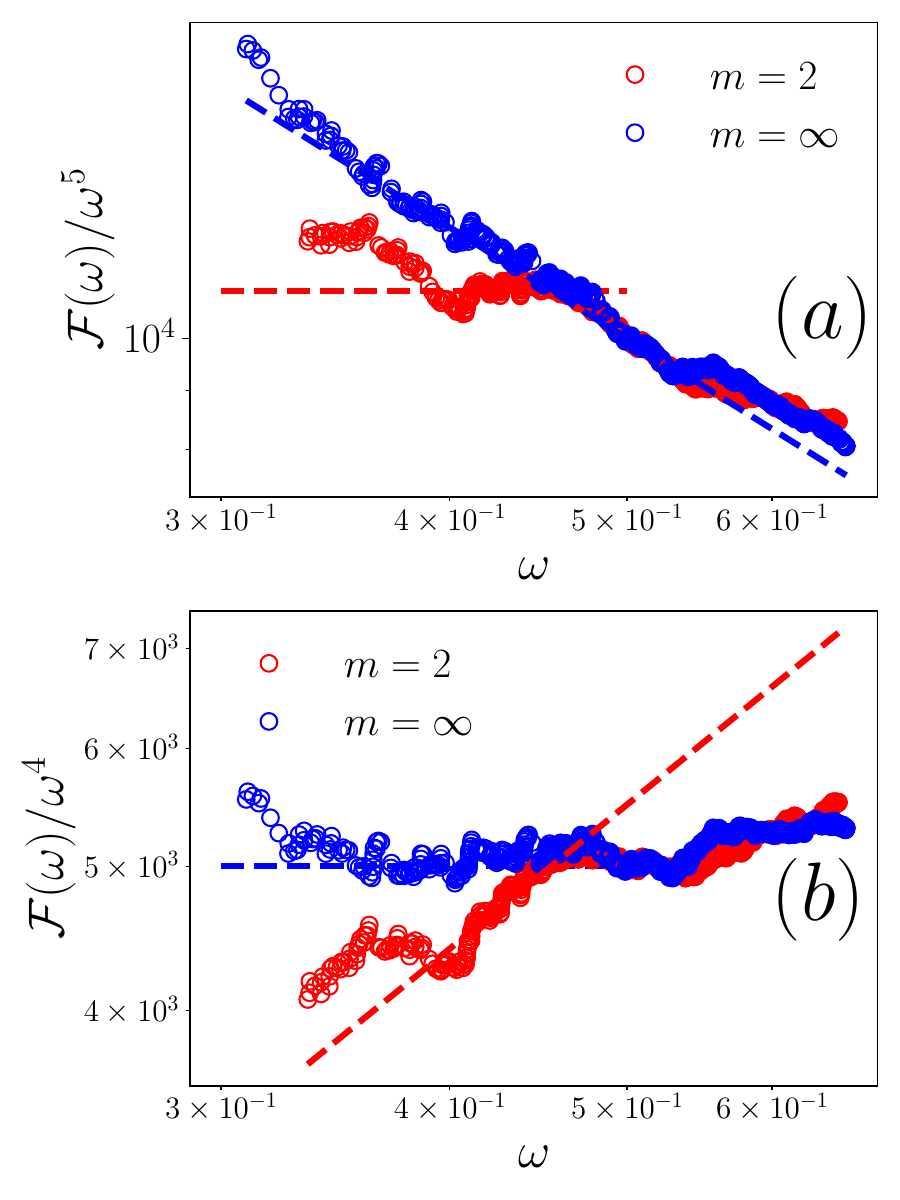}
	\caption{Panel (a): plot of the function $F(\omega)/\omega^5$ for $m$=2 (red dots) and $m$=$\infty$ (blue dots). The horizontal dashed red line emphasizes the expected $s$=4 for $m$=2. The blue dashed line has a slope equal to -1 (i.e. $s=3$). Panel (b): plot of the function $F(\omega)/\omega^4$ for $m$=2 (red dots) and $m$=$\infty$ (blue dots). The horizontal dashed blue line emphasizes the expected $s$=3 for $m$=$\infty$. The red dashed line has a slope equal to 1 (i.e. $s=4$)}
\label{fig1}
\end{figure}

(iv)
In our paper \cite{schirmacher24} we point out that the 
type-II excitations in systems, with pairwise interaction
potentials with a minimum, generically should have a contribution
to the DoS, scaling as $s$ = 5. We quoted simulations
\cite{krishnan22,krishnan23}, in which Lennard-Jones potentials
are used and in which $s$ = 5 is observed. We did not perform any
simulations with such potentials ourselves and only want to point out
that care must be taken to avoid artifacts associated with the
tapering. As the tapering-induced terms in the DoS scale with a
lower $s$, there will be a crossover between the two contributions.
We leave the calculations presented in Ref. \cite{lerner25}
to refute our statement
uncommented, as we cannot retrace the details of these calculations,
especially the  sample quenching procedure.
Of course, we cannot exclude that the theory presented in~\cite{schirmacher24} fails for systems with 
attractive interactions for some unknown reason.

\begin{figure}[t]
\includegraphics[width=.95\columnwidth]{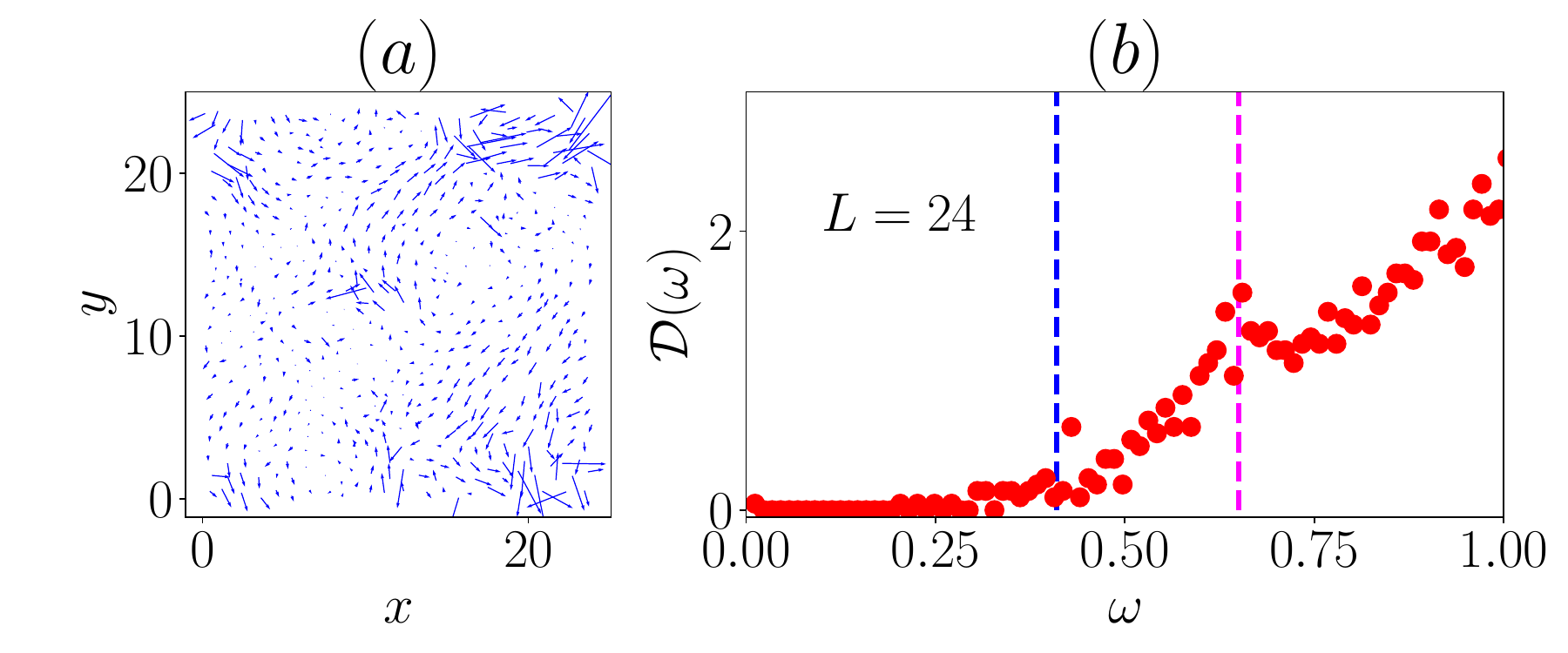}
\caption{Panel (a) reports the sketch of a mode at $\omega \approx$ 0.4 (blue dashed line in panel (b)) which show a vortex-like feature. This mode is at lower frequency with respect to the lowest transverse phonons resonance (magenta dashed line in panel (b)) found at $\omega \approx$ 0.65.}
\label{fig2}
\end{figure}

{\it (v)}
In our opinion, the most important result of our paper 
\cite{schirmacher24}
points to the existence of vortex-like modes, being related
to local frozen-in stresses.
The authors of Ref. \cite{lerner25} state that the vortex-like modes are a superposition of transverse standing waves (phonons). To show this, they report in \cite[Fig.~3]{lerner25} the eigenvectors of a couple of transverse modes with ``identical wavelength", and their superposition which resembles a vortex-like mode. They conclude that {\it all} the vortex-like modes are of this kind. To support our statements, in Fig.~2 we report a vortex-like mode, together with the DoS of the system from where the mode has been extracted. One can easily see that the frequency of the selected mode
(blue dashed line, $\omega \approx$ 0.4)
is by far lower than that of the lowest transverse phonons (magenta dashed line, $\omega \approx$ 0.65). 
Thus this low-frequency mode is genuine and does not originate from the superposition of waves. Furthermore, and more importantly, because of level repulsion acting for extended modes in disordered systems, two phonon modes {\it cannot have} the same frequency, and therefore cannot be combined to generate a superposition. Further clarification on this point would be needed to better understand the procedure used to generate~\cite[Fig.3]{lerner25}.

\new{\section{Discussion}}

Our \new{new} data 
provide an example
of a system, which clearly shows a sensitive
dependence of the DoS exponent $s$ on the tapering procedure,
which points to the existence
of stress-related vortex-like modes. 
We have demonstrated that \new{the frequency scaling of the DoS, calculated from
molecular-dynamics, is sensitive to the cutoff-smoothing (tapering) procedure,
and that therefore}
the $\omega^4$ dependence of the DoS \sout{is not} \new{may not be} universal.
\new{We believe that the arguments presented 
in the 
letter of Lerner and Bouchbinder \cite{lerner25} are
not in contradiction with the results of \cite{schirmacher24}}.

In future simulations care must be taken 
to investigate the dependence of the
data on the employed algorithms, such as the
degree of the cutoff-smoothing (tapering) of the potential.
\sk
\acknowledgments

WS thanks E. Lerner for helpful discussions.
DK acknowledges support from the Foundation for Polish Science in Poland through the FENG.02.02-IP.05-0177/23 project. This work was carried out in part within the ``Projektowanie Ulepszonych Szkie\l Metalicznych" project (FENG.02.02-IP.05-0177/23) under the 2.2 First Team programme of the Foundation for Polish Science co-financed by the European Union from the European Funds for Smart Economy 2021-2027 (FENG).


\begin{thebibliography}{37}%
\makeatletter
\providecommand \@ifxundefined [1]{%
 \@ifx{#1\undefined}
}%
\providecommand \@ifnum [1]{%
 \ifnum #1\expandafter \@firstoftwo
 \else \expandafter \@secondoftwo
 \fi
}%
\providecommand \@ifx [1]{%
 \ifx #1\expandafter \@firstoftwo
 \else \expandafter \@secondoftwo
 \fi
}%
\providecommand \natexlab [1]{#1}%
\providecommand \enquote  [1]{``#1''}%
\providecommand \bibnamefont  [1]{#1}%
\providecommand \bibfnamefont [1]{#1}%
\providecommand \citenamefont [1]{#1}%
\providecommand \href@noop [0]{\@secondoftwo}%
\providecommand \href [0]{\begingroup \@sanitize@url \@href}%
\providecommand \@href[1]{\@@startlink{#1}\@@href}%
\providecommand \@@href[1]{\endgroup#1\@@endlink}%
\providecommand \@sanitize@url [0]{\catcode `\\12\catcode `\$12\catcode
  `\&12\catcode `\#12\catcode `\^12\catcode `\_12\catcode `\%12\relax}%
\providecommand \@@startlink[1]{}%
\providecommand \@@endlink[0]{}%
\providecommand \url  [0]{\begingroup\@sanitize@url \@url }%
\providecommand \@url [1]{\endgroup\@href {#1}{\urlprefix }}%
\providecommand \urlprefix  [0]{URL }%
\providecommand \Eprint [0]{\href }%
\providecommand \doibase [0]{https://doi.org/}%
\providecommand \selectlanguage [0]{\@gobble}%
\providecommand \bibinfo  [0]{\@secondoftwo}%
\providecommand \bibfield  [0]{\@secondoftwo}%
\providecommand \translation [1]{[#1]}%
\providecommand \BibitemOpen [0]{}%
\providecommand \bibitemStop [0]{}%
\providecommand \bibitemNoStop [0]{.\EOS\space}%
\providecommand \EOS [0]{\spacefactor3000\relax}%
\providecommand \BibitemShut  [1]{\csname bibitem#1\endcsname}%
\let\auto@bib@innerbib\@empty
\bibitem [{\citenamefont {Elliott}(1984)}]{elliott84}%
  \BibitemOpen
  \bibfield  {author} {\bibinfo {author} {\bibfnamefont {S.~R.}\ \bibnamefont
  {Elliott}},\ }\href@noop {} {\emph {\bibinfo {title} {The Physics of
  Amorphous Materials}}}\ (\bibinfo  {publisher} {Longman},\ \bibinfo {address}
  {New York},\ \bibinfo {year} {1984})\BibitemShut {NoStop}%
\bibitem [{\citenamefont {Binder}\ and\ \citenamefont {Kob}(2005)}]{binder05}%
  \BibitemOpen
  \bibfield  {author} {\bibinfo {author} {\bibfnamefont {K.}~\bibnamefont
  {Binder}}\ and\ \bibinfo {author} {\bibfnamefont {W.}~\bibnamefont {Kob}},\
  }\href@noop {} {\emph {\bibinfo {title} {Glassy Materials and Disordered
  Solids: An Introduction to Their Statistical Mechanics}}}\ (\bibinfo
  {publisher} {World Scientific},\ \bibinfo {address} {Singapore},\ \bibinfo
  {year} {2005})\BibitemShut {NoStop}%
\bibitem [{\citenamefont {Ramos}(2022)}]{ramos22}%
  \BibitemOpen
  \bibfield  {author} {\bibinfo {author} {\bibfnamefont {M.~A.}\ \bibnamefont
  {Ramos}},\ }\href@noop {} {\emph {\bibinfo {title} {Low-Temperature Thermal
  and Vibrational Properties of Disoredered Solids (A Half-Century of universal
  "anomalies" of glasses)}}}\ (\bibinfo  {publisher} {World Scientific},\
  \bibinfo {address} {New Jersey},\ \bibinfo {year} {2022})\BibitemShut
  {NoStop}%
\bibitem [{\citenamefont {Berne}\ and\ \citenamefont {Pecora}(1976)}]{berne76}%
  \BibitemOpen
  \bibfield  {author} {\bibinfo {author} {\bibfnamefont {B.~J.}\ \bibnamefont
  {Berne}}\ and\ \bibinfo {author} {\bibfnamefont {R.}~\bibnamefont {Pecora}},\
  }\href@noop {} {\emph {\bibinfo {title} {Dynamic light scattering}}}\
  (\bibinfo  {publisher} {Wiley},\ \bibinfo {address} {New York},\ \bibinfo
  {year} {1976})\BibitemShut {NoStop}%
\bibitem [{\citenamefont {Schober}(2014)}]{schober14}%
  \BibitemOpen
  \bibfield  {author} {\bibinfo {author} {\bibfnamefont {H.}~\bibnamefont
  {Schober}},\ }\bibfield  {title} {\bibinfo {title} {An introduction to the
  theory of nuclear neutron scattering in condensed matter},\ }\href@noop {}
  {\bibfield  {journal} {\bibinfo  {journal} {J. Neutron Res.}\ }\textbf
  {\bibinfo {volume} {17}},\ \bibinfo {pages} {109} (\bibinfo {year}
  {2014})}\BibitemShut {NoStop}%
\bibitem [{\citenamefont {Baron}(2020)}]{baron20}%
  \BibitemOpen
  \bibfield  {author} {\bibinfo {author} {\bibfnamefont {A.~Q.~R.}\
  \bibnamefont {Baron}},\ }\bibfield  {title} {\bibinfo {title} {Introduction
  to high-resolution inelastic x-ray scattering},\ }in\ \href@noop {} {\emph
  {\bibinfo {booktitle} {Synchrotron light sources \& free electron lasers}}},\
  \bibinfo {editor} {edited by\ \bibinfo {editor} {\bibfnamefont {E.~J.}\
  \bibnamefont {{\it et al.}}}}\ (\bibinfo {year} {2020})\ p.\ \bibinfo {pages}
  {1643}\BibitemShut {NoStop}%
\bibitem [{Note1()}]{Note1}%
  \BibitemOpen
  \bibinfo {note} {See \cite {ramos22,schirmacher24b} for an extensive
  discussion of the boson peak.}\BibitemShut {Stop}%
\bibitem [{\citenamefont {Schirmacher}(2006)}]{schirm06}%
  \BibitemOpen
  \bibfield  {author} {\bibinfo {author} {\bibfnamefont {W.}~\bibnamefont
  {Schirmacher}},\ }\bibfield  {title} {\bibinfo {title} {{Thermal conductivity
  of glassy materials and the ``boson peak''}},\ }\href@noop {} {\bibfield
  {journal} {\bibinfo  {journal} {Europhys. Lett.}\ }\textbf {\bibinfo {volume}
  {73}},\ \bibinfo {pages} {892} (\bibinfo {year} {2006})}\BibitemShut
  {NoStop}%
\bibitem [{\citenamefont {Schirmacher}\ \emph {et~al.}(2007)\citenamefont
  {Schirmacher}, \citenamefont {G.Ruocco},\ and\ \citenamefont
  {Scopigno}}]{schirm07}%
  \BibitemOpen
  \bibfield  {author} {\bibinfo {author} {\bibfnamefont {W.}~\bibnamefont
  {Schirmacher}}, \bibinfo {author} {\bibnamefont {G.Ruocco}},\ and\ \bibinfo
  {author} {\bibfnamefont {T.}~\bibnamefont {Scopigno}},\ }\bibfield  {title}
  {\bibinfo {title} {{Acoustic attenuation in glasses and its relation with the
  Boson Peak}},\ }\href@noop {} {\bibfield  {journal} {\bibinfo  {journal}
  {Phys. Rev. Lett.}\ }\textbf {\bibinfo {volume} {98}},\ \bibinfo {pages}
  {025501} (\bibinfo {year} {2007})}\BibitemShut {NoStop}%
\bibitem [{\citenamefont {Marruzzo}\ \emph {et~al.}(2013)\citenamefont
  {Marruzzo}, \citenamefont {Schirmacher}, \citenamefont {Fratalocchi},\ and\
  \citenamefont {Ruocco}}]{marruzzo13}%
  \BibitemOpen
  \bibfield  {author} {\bibinfo {author} {\bibfnamefont {A.}~\bibnamefont
  {Marruzzo}}, \bibinfo {author} {\bibfnamefont {W.}~\bibnamefont
  {Schirmacher}}, \bibinfo {author} {\bibfnamefont {A.}~\bibnamefont
  {Fratalocchi}},\ and\ \bibinfo {author} {\bibfnamefont {G.}~\bibnamefont
  {Ruocco}},\ }\bibfield  {title} {\bibinfo {title} {Heterogeneous shear
  elasticity of glasses: the origin of the boson peak},\ }\href@noop {}
  {\bibfield  {journal} {\bibinfo  {journal} {Sci. Rep.}\ }\textbf {\bibinfo
  {volume} {3}},\ \bibinfo {pages} {1} (\bibinfo {year} {2013})}\BibitemShut
  {NoStop}%
\bibitem [{\citenamefont {K\"ohler}\ \emph {et~al.}(2013)\citenamefont
  {K\"ohler}, \citenamefont {Ruocco},\ and\ \citenamefont
  {Schirmacher}}]{kohler13}%
  \BibitemOpen
  \bibfield  {author} {\bibinfo {author} {\bibfnamefont {S.}~\bibnamefont
  {K\"ohler}}, \bibinfo {author} {\bibfnamefont {R.}~\bibnamefont {Ruocco}},\
  and\ \bibinfo {author} {\bibfnamefont {W.}~\bibnamefont {Schirmacher}},\
  }\bibfield  {title} {\bibinfo {title} {Coherent-potential approximation for
  diffusion and wave propagation in topologically disordered systems},\
  }\href@noop {} {\bibfield  {journal} {\bibinfo  {journal} {Phys. Rev. B}\
  }\textbf {\bibinfo {volume} {88}},\ \bibinfo {pages} {064203} (\bibinfo
  {year} {2013})}\BibitemShut {NoStop}%
\bibitem [{\citenamefont {Schirmacher}\ \emph {et~al.}(2014)\citenamefont
  {Schirmacher}, \citenamefont {Scopigno},\ and\ \citenamefont
  {Ruocco}}]{schirm14}%
  \BibitemOpen
  \bibfield  {author} {\bibinfo {author} {\bibfnamefont {W.}~\bibnamefont
  {Schirmacher}}, \bibinfo {author} {\bibfnamefont {T.}~\bibnamefont
  {Scopigno}},\ and\ \bibinfo {author} {\bibfnamefont {G.}~\bibnamefont
  {Ruocco}},\ }\bibfield  {title} {\bibinfo {title} {Theory of vibrational
  anomalies in glasses},\ }\href@noop {} {\bibfield  {journal} {\bibinfo
  {journal} {J. Noncryst. Sol.}\ }\textbf {\bibinfo {volume} {407}},\ \bibinfo
  {pages} {133} (\bibinfo {year} {2014})}\BibitemShut {NoStop}%
\bibitem [{\citenamefont {Lerner}\ \emph {et~al.}(2016)\citenamefont {Lerner},
  \citenamefont {D\"uring},\ and\ \citenamefont {Bouchbinder}}]{Lerner2016}%
  \BibitemOpen
  \bibfield  {author} {\bibinfo {author} {\bibfnamefont {E.}~\bibnamefont
  {Lerner}}, \bibinfo {author} {\bibfnamefont {G.}~\bibnamefont {D\"uring}},\
  and\ \bibinfo {author} {\bibfnamefont {E.}~\bibnamefont {Bouchbinder}},\
  }\bibfield  {title} {\bibinfo {title} {Statistics and properties of
  low-frequency vibrational modes in structural glasses},\ }\href
  {https://doi.org/10.1103/PhysRevLett.117.035501} {\bibfield  {journal}
  {\bibinfo  {journal} {Phys. Rev. Lett.}\ }\textbf {\bibinfo {volume} {117}},\
  \bibinfo {pages} {035501} (\bibinfo {year} {2016})}\BibitemShut {NoStop}%
\bibitem [{\citenamefont {Lerner}\ and\ \citenamefont
  {Bouchbinder}(2017)}]{Lerner2017}%
  \BibitemOpen
  \bibfield  {author} {\bibinfo {author} {\bibfnamefont {E.}~\bibnamefont
  {Lerner}}\ and\ \bibinfo {author} {\bibfnamefont {E.}~\bibnamefont
  {Bouchbinder}},\ }\bibfield  {title} {\bibinfo {title} {Effect of
  instantaneous and continuous quenches on the density of vibrational modes in
  model glasses},\ }\href@noop {} {\bibfield  {journal} {\bibinfo  {journal}
  {Phys. Rev. E}\ }\textbf {\bibinfo {volume} {96}},\ \bibinfo {pages}
  {020104(R)} (\bibinfo {year} {2017})}\BibitemShut {NoStop}%
\bibitem [{\citenamefont {Angelani}\ \emph {et~al.}(2018)\citenamefont
  {Angelani}, \citenamefont {Paoluzzi}, \citenamefont {Parisi},\ and\
  \citenamefont {Ruocco}}]{parisi18}%
  \BibitemOpen
  \bibfield  {author} {\bibinfo {author} {\bibfnamefont {L.}~\bibnamefont
  {Angelani}}, \bibinfo {author} {\bibfnamefont {M.}~\bibnamefont {Paoluzzi}},
  \bibinfo {author} {\bibfnamefont {G.}~\bibnamefont {Parisi}},\ and\ \bibinfo
  {author} {\bibfnamefont {G.}~\bibnamefont {Ruocco}},\ }\bibfield  {title}
  {\bibinfo {title} {Probing the non-debye low-frequency excitations in glasses
  through random pinning},\ }\href@noop {} {\bibfield  {journal} {\bibinfo
  {journal} {Proc. Nat. Acad. Sci.}\ }\textbf {\bibinfo {volume} {115}},\
  \bibinfo {pages} {8700} (\bibinfo {year} {2018})}\BibitemShut {NoStop}%
\bibitem [{\citenamefont {Paoluzzi}\ \emph {et~al.}(2019)\citenamefont
  {Paoluzzi}, \citenamefont {Angelani}, \citenamefont {Parisi},\ and\
  \citenamefont {Ruocco}}]{parisi19}%
  \BibitemOpen
  \bibfield  {author} {\bibinfo {author} {\bibfnamefont {M.}~\bibnamefont
  {Paoluzzi}}, \bibinfo {author} {\bibfnamefont {L.}~\bibnamefont {Angelani}},
  \bibinfo {author} {\bibfnamefont {G.}~\bibnamefont {Parisi}},\ and\ \bibinfo
  {author} {\bibfnamefont {G.}~\bibnamefont {Ruocco}},\ }\bibfield  {title}
  {\bibinfo {title} {Relatiion between heterogeneous frozen regions in
  supercooled liquids and non-debye spectrum in the corresponding glasses},\
  }\href@noop {} {\bibfield  {journal} {\bibinfo  {journal} {Phys. Rev. Lett}\
  }\textbf {\bibinfo {volume} {123}},\ \bibinfo {pages} {155502} (\bibinfo
  {year} {2019})}\BibitemShut {NoStop}%
\bibitem [{\citenamefont {Lerner}\ and\ \citenamefont
  {Bouchbinder}(2021)}]{lernerbouchbinder21}%
  \BibitemOpen
  \bibfield  {author} {\bibinfo {author} {\bibfnamefont {E.}~\bibnamefont
  {Lerner}}\ and\ \bibinfo {author} {\bibfnamefont {E.}~\bibnamefont
  {Bouchbinder}},\ }\bibfield  {title} {\bibinfo {title} {Low-energy
  quasilocalized excitations in structural glasses},\ }\href@noop {} {\bibfield
   {journal} {\bibinfo  {journal} {J. Chem. Phys.}\ }\textbf {\bibinfo {volume}
  {155}},\ \bibinfo {pages} {200901} (\bibinfo {year} {2021})}\BibitemShut
  {NoStop}%
\bibitem [{\citenamefont {Paoluzzi}\ \emph {et~al.}(2021)\citenamefont
  {Paoluzzi}, \citenamefont {Angelani}, \citenamefont {Parisi},\ and\
  \citenamefont {Ruocco}}]{parisi21}%
  \BibitemOpen
  \bibfield  {author} {\bibinfo {author} {\bibfnamefont {M.}~\bibnamefont
  {Paoluzzi}}, \bibinfo {author} {\bibfnamefont {L.}~\bibnamefont {Angelani}},
  \bibinfo {author} {\bibfnamefont {G.}~\bibnamefont {Parisi}},\ and\ \bibinfo
  {author} {\bibfnamefont {G.}~\bibnamefont {Ruocco}},\ }\bibfield  {title}
  {\bibinfo {title} {Probing the debye spectrum in glasses using small system
  sizes},\ }\href@noop {} {\bibfield  {journal} {\bibinfo  {journal} {Phys.
  Rev. Res.}\ }\textbf {\bibinfo {volume} {2}},\ \bibinfo {pages} {043248}
  (\bibinfo {year} {2021})}\BibitemShut {NoStop}%
\bibitem [{\citenamefont {Schirmacher}\ \emph {et~al.}(2024)\citenamefont
  {Schirmacher}, \citenamefont {Paoluzzi}, \citenamefont {Mocanu},
  \citenamefont {Khomenko}, \citenamefont {Szamel}, \citenamefont {Zamponi},\
  and\ \citenamefont {Ruocco}}]{schirmacher24}%
  \BibitemOpen
  \bibfield  {author} {\bibinfo {author} {\bibfnamefont {W.}~\bibnamefont
  {Schirmacher}}, \bibinfo {author} {\bibfnamefont {M.}~\bibnamefont
  {Paoluzzi}}, \bibinfo {author} {\bibfnamefont {F.~C.}\ \bibnamefont
  {Mocanu}}, \bibinfo {author} {\bibfnamefont {D.}~\bibnamefont {Khomenko}},
  \bibinfo {author} {\bibfnamefont {G.}~\bibnamefont {Szamel}}, \bibinfo
  {author} {\bibfnamefont {F.}~\bibnamefont {Zamponi}},\ and\ \bibinfo {author}
  {\bibfnamefont {G.}~\bibnamefont {Ruocco}},\ }\bibfield  {title} {\bibinfo
  {title} {The nature of non-phononic excitations in disordered systems},\
  }\href@noop {} {\bibfield  {journal} {\bibinfo  {journal} {Nature
  Communications}\ }\textbf {\bibinfo {volume} {15}},\ \bibinfo {pages} {3107}
  (\bibinfo {year} {2024})}\BibitemShut {NoStop}%
\bibitem [{\citenamefont {Lerner}\ and\ \citenamefont
  {Bouchbinder}(2025)}]{lerner25}%
  \BibitemOpen
  \bibfield  {author} {\bibinfo {author} {\bibfnamefont {E.}~\bibnamefont
  {Lerner}}\ and\ \bibinfo {author} {\bibfnamefont {E.}~\bibnamefont
  {Bouchbinder}},\ }\bibfield  {title} {\bibinfo {title} {Testing the
  heterogeneous-elasticity theory for low-energy excitations in structural
  glasses},\ }\href@noop {} {\bibfield  {journal} {\bibinfo  {journal} {Phys.
  Rev. E}\ }\textbf {\bibinfo {volume} {111}},\ \bibinfo {pages} {L013402}
  (\bibinfo {year} {2025})}\BibitemShut {NoStop}%
\bibitem [{\citenamefont {Ashcroft}\ and\ \citenamefont
  {Mermin}(1976)}]{ashcroft76a}%
  \BibitemOpen
  \bibfield  {author} {\bibinfo {author} {\bibfnamefont {N.~W.}\ \bibnamefont
  {Ashcroft}}\ and\ \bibinfo {author} {\bibfnamefont {D.}~\bibnamefont
  {Mermin}},\ }\href@noop {} {\emph {\bibinfo {title} {Solid state physics}}}\
  (\bibinfo  {publisher} {Harcourt College Publishers},\ \bibinfo {address}
  {Fort Worth, USA},\ \bibinfo {year} {1976})\ p.\ \bibinfo {pages}
  {443}\BibitemShut {NoStop}%
\bibitem [{\citenamefont {Lutsko}(1988)}]{lutsko88}%
  \BibitemOpen
  \bibfield  {author} {\bibinfo {author} {\bibfnamefont {J.~F.}\ \bibnamefont
  {Lutsko}},\ }\bibfield  {title} {\bibinfo {title} {Stress and elastic
  constants in anisotropic solids: Molecular dynamics techniques},\ }\href@noop
  {} {\bibfield  {journal} {\bibinfo  {journal} {J. Appl. Phys.}\ }\textbf
  {\bibinfo {volume} {64}},\ \bibinfo {pages} {1152} (\bibinfo {year}
  {1988})}\BibitemShut {NoStop}%
\bibitem [{\citenamefont {Lutsko}(1989)}]{lutsko89}%
  \BibitemOpen
  \bibfield  {author} {\bibinfo {author} {\bibfnamefont {J.~F.}\ \bibnamefont
  {Lutsko}},\ }\bibfield  {title} {\bibinfo {title} {Generalized expressions
  for the calculation of eiastic constants by computer simulation},\
  }\href@noop {} {\bibfield  {journal} {\bibinfo  {journal} {J. Appl. Phys.}\
  }\textbf {\bibinfo {volume} {65}},\ \bibinfo {pages} {2991} (\bibinfo {year}
  {1989})}\BibitemShut {NoStop}%
\bibitem [{\citenamefont {Alexander}(1998)}]{alexander98}%
  \BibitemOpen
  \bibfield  {author} {\bibinfo {author} {\bibfnamefont {S.}~\bibnamefont
  {Alexander}},\ }\bibfield  {title} {\bibinfo {title} {Amorphous solids: their
  structure, lattice dynamics and elasticity},\ }\href@noop {} {\bibfield
  {journal} {\bibinfo  {journal} {Phys. Reports}\ }\textbf {\bibinfo {volume}
  {296}},\ \bibinfo {pages} {65} (\bibinfo {year} {1998})}\BibitemShut
  {NoStop}%
\bibitem [{\citenamefont {Alexander}(1984)}]{alexander84}%
  \BibitemOpen
  \bibfield  {author} {\bibinfo {author} {\bibfnamefont {S.}~\bibnamefont
  {Alexander}},\ }\bibfield  {title} {\bibinfo {title} {Is the elastic energy
  of amorphous materials rotationally invariant?},\ }\href@noop {} {\bibfield
  {journal} {\bibinfo  {journal} {J. Physique}\ }\textbf {\bibinfo {volume}
  {45}},\ \bibinfo {pages} {1939} (\bibinfo {year} {1984})}\BibitemShut
  {NoStop}%
\bibitem [{\citenamefont {Mehta}(1967)}]{mehta67}%
  \BibitemOpen
  \bibfield  {author} {\bibinfo {author} {\bibfnamefont {M.~L.}\ \bibnamefont
  {Mehta}},\ }\href@noop {} {\emph {\bibinfo {title} {Random Matrices}}}\
  (\bibinfo  {publisher} {Academic Press},\ \bibinfo {address} {New York},\
  \bibinfo {year} {1967})\BibitemShut {NoStop}%
\bibitem [{\citenamefont {Schirmacher}\ \emph {et~al.}(1998)\citenamefont
  {Schirmacher}, \citenamefont {Diezemann},\ and\ \citenamefont
  {Ganter}}]{schirm98}%
  \BibitemOpen
  \bibfield  {author} {\bibinfo {author} {\bibfnamefont {W.}~\bibnamefont
  {Schirmacher}}, \bibinfo {author} {\bibfnamefont {G.}~\bibnamefont
  {Diezemann}},\ and\ \bibinfo {author} {\bibfnamefont {C.}~\bibnamefont
  {Ganter}},\ }\bibfield  {title} {\bibinfo {title} {Harmonic vibrational
  excitations in disordered solids and the ``boson peak''},\ }\href@noop {}
  {\bibfield  {journal} {\bibinfo  {journal} {Phys. Rev. Lett.}\ }\textbf
  {\bibinfo {volume} {81}},\ \bibinfo {pages} {136} (\bibinfo {year}
  {1998})}\BibitemShut {NoStop}%
\bibitem [{\citenamefont {Allen}\ \emph {et~al.}(1999)\citenamefont {Allen},
  \citenamefont {Feldman},\ and\ \citenamefont {Fabian}}]{allen99}%
  \BibitemOpen
  \bibfield  {author} {\bibinfo {author} {\bibfnamefont {P.~B.}\ \bibnamefont
  {Allen}}, \bibinfo {author} {\bibfnamefont {J.~L.}\ \bibnamefont {Feldman}},\
  and\ \bibinfo {author} {\bibfnamefont {J.}~\bibnamefont {Fabian}},\
  }\bibfield  {title} {\bibinfo {title} {Diffusons, locons and propagons:
  character of atomic vibrations in amorphous si},\ }\href@noop {} {\bibfield
  {journal} {\bibinfo  {journal} {Philos. Mag.}\ }\textbf {\bibinfo {volume}
  {79}},\ \bibinfo {pages} {1715} (\bibinfo {year} {1999})}\BibitemShut
  {NoStop}%
\bibitem [{Note2()}]{Note2}%
  \BibitemOpen
  \bibinfo {note} {The stresses $\sigma _\ell $ may take both signs as a
  consequence of the external pressure imposed by the boundary conditions, see
  \cite {schirmacher24}.}\BibitemShut {Stop}%
\bibitem [{\citenamefont {Franz}\ \emph {et~al.}(2015)\citenamefont {Franz},
  \citenamefont {Parisi}, \citenamefont {Urbani},\ and\ \citenamefont
  {Zamponi}}]{franz15}%
  \BibitemOpen
  \bibfield  {author} {\bibinfo {author} {\bibfnamefont {S.}~\bibnamefont
  {Franz}}, \bibinfo {author} {\bibfnamefont {G.}~\bibnamefont {Parisi}},
  \bibinfo {author} {\bibfnamefont {P.}~\bibnamefont {Urbani}},\ and\ \bibinfo
  {author} {\bibfnamefont {F.}~\bibnamefont {Zamponi}},\ }\bibfield  {title}
  {\bibinfo {title} {Universal spectrum of normal modes in low-temperature
  glasses},\ }\href@noop {} {\bibfield  {journal} {\bibinfo  {journal} {Proc.
  Nat. Acad. Sci.}\ }\textbf {\bibinfo {volume} {112}},\ \bibinfo {pages}
  {14539} (\bibinfo {year} {2015})}\BibitemShut {NoStop}%
\bibitem [{\citenamefont {Wang}\ \emph {et~al.}(2021)\citenamefont {Wang},
  \citenamefont {Szamel},\ and\ \citenamefont {Flenner}}]{wang21}%
  \BibitemOpen
  \bibfield  {author} {\bibinfo {author} {\bibfnamefont {L.}~\bibnamefont
  {Wang}}, \bibinfo {author} {\bibfnamefont {G.}~\bibnamefont {Szamel}},\ and\
  \bibinfo {author} {\bibfnamefont {E.}~\bibnamefont {Flenner}},\ }\bibfield
  {title} {\bibinfo {title} {Low-frequency excess vibrational modes in
  two-dimensional glasses},\ }\href@noop {} {\bibfield  {journal} {\bibinfo
  {journal} {Phys. Rev. Lett.}\ }\textbf {\bibinfo {volume} {127}},\ \bibinfo
  {pages} {248001} (\bibinfo {year} {2021})}\BibitemShut {NoStop}%
\bibitem [{\citenamefont {Franz}\ \emph {et~al.}(2025)\citenamefont {Franz},
  \citenamefont {Lupo}, \citenamefont {Nicoletti}, \citenamefont {Parisi},\
  and\ \citenamefont {Ricci-Tersenghi}}]{franz25}%
  \BibitemOpen
  \bibfield  {author} {\bibinfo {author} {\bibfnamefont {S.}~\bibnamefont
  {Franz}}, \bibinfo {author} {\bibfnamefont {C.}~\bibnamefont {Lupo}},
  \bibinfo {author} {\bibfnamefont {F.}~\bibnamefont {Nicoletti}}, \bibinfo
  {author} {\bibfnamefont {G.}~\bibnamefont {Parisi}},\ and\ \bibinfo {author}
  {\bibfnamefont {F.}~\bibnamefont {Ricci-Tersenghi}},\ }\bibfield  {title}
  {\bibinfo {title} {Soft modes in vector spin glass models on sparse random
  graphs},\ }\href@noop {} {\bibfield  {journal} {\bibinfo  {journal} {Phys.
  Rev. B}\ }\textbf {\bibinfo {volume} {111}},\ \bibinfo {pages} {014203}
  (\bibinfo {year} {2025})}\BibitemShut {NoStop}%
\bibitem [{\citenamefont {Schober}\ and\ \citenamefont
  {Oligschleger}(1996)}]{schober96}%
  \BibitemOpen
  \bibfield  {author} {\bibinfo {author} {\bibfnamefont {H.}~\bibnamefont
  {Schober}}\ and\ \bibinfo {author} {\bibfnamefont {C.}~\bibnamefont
  {Oligschleger}},\ }\bibfield  {title} {\bibinfo {title} {Low-frequency
  vibrations in a model glass},\ }\href@noop {} {\bibfield  {journal} {\bibinfo
   {journal} {Phys. Rev. B}\ }\textbf {\bibinfo {volume} {53}},\ \bibinfo
  {pages} {11469} (\bibinfo {year} {1996})}\BibitemShut {NoStop}%
\bibitem [{\citenamefont {Schober}\ and\ \citenamefont
  {Ruocco}(2004)}]{schober04}%
  \BibitemOpen
  \bibfield  {author} {\bibinfo {author} {\bibfnamefont {H.}~\bibnamefont
  {Schober}}\ and\ \bibinfo {author} {\bibfnamefont {G.}~\bibnamefont
  {Ruocco}},\ }\bibfield  {title} {\bibinfo {title} {Size effects and
  quasilocalized vibrations},\ }\href@noop {} {\bibfield  {journal} {\bibinfo
  {journal} {Philos. Magazine}\ }\textbf {\bibinfo {volume} {84}},\ \bibinfo
  {pages} {1361} (\bibinfo {year} {2004})}\BibitemShut {NoStop}%
\bibitem [{\citenamefont {Krishnan}\ \emph {et~al.}(2022)\citenamefont
  {Krishnan}, \citenamefont {Ramola},\ and\ \citenamefont
  {Karmakar}}]{krishnan22}%
  \BibitemOpen
  \bibfield  {author} {\bibinfo {author} {\bibfnamefont {V.~V.}\ \bibnamefont
  {Krishnan}}, \bibinfo {author} {\bibfnamefont {K.}~\bibnamefont {Ramola}},\
  and\ \bibinfo {author} {\bibfnamefont {S.}~\bibnamefont {Karmakar}},\
  }\bibfield  {title} {\bibinfo {title} {Universal non-debye low-frequency
  vibrations in sheared amorphous solids},\ }\href@noop {} {\bibfield
  {journal} {\bibinfo  {journal} {Soft Matter}\ }\textbf {\bibinfo {volume}
  {18}},\ \bibinfo {pages} {3395} (\bibinfo {year} {2022})}\BibitemShut
  {NoStop}%
\bibitem [{\citenamefont {Chakraborty}\ \emph {et~al.}(2022)\citenamefont
  {Chakraborty}, \citenamefont {Krishnan}, \citenamefont {Ramola},\ and\
  \citenamefont {Karmakar}}]{krishnan23}%
  \BibitemOpen
  \bibfield  {author} {\bibinfo {author} {\bibfnamefont {S.}~\bibnamefont
  {Chakraborty}}, \bibinfo {author} {\bibfnamefont {V.~V.}\ \bibnamefont
  {Krishnan}}, \bibinfo {author} {\bibfnamefont {K.}~\bibnamefont {Ramola}},\
  and\ \bibinfo {author} {\bibfnamefont {S.}~\bibnamefont {Karmakar}},\
  }\bibfield  {title} {\bibinfo {title} {Enhanced vibrational stability in
  glass droplets},\ }\bibfield  {journal} {\bibinfo  {journal} {PNAS Nexus}\
  }\href {https://doi.org/https://doi.org/10.1093/pnasnexus/pgrad289}
  {https://doi.org/10.1093/pnasnexus/pgrad289} (\bibinfo {year}
  {2022})\BibitemShut {NoStop}%
\bibitem [{\citenamefont {Schirmacher}\ and\ \citenamefont
  {Ruocco}(2024)}]{schirmacher24b}%
  \BibitemOpen
  \bibfield  {author} {\bibinfo {author} {\bibfnamefont {W.}~\bibnamefont
  {Schirmacher}}\ and\ \bibinfo {author} {\bibfnamefont {G.}~\bibnamefont
  {Ruocco}},\ }\bibfield  {title} {\bibinfo {title} {Vibrational excitations in
  disordered solids},\ }in\ \href@noop {} {\emph {\bibinfo {booktitle}
  {Eccyclopedia of condensed matter physics 2e}}},\ Vol.~\bibinfo {volume}
  {5},\ \bibinfo {editor} {edited by\ \bibinfo {editor} {\bibfnamefont
  {T.}~\bibnamefont {Chakraborty}}}\ (\bibinfo  {publisher} {Elsevier},\
  \bibinfo {address} {Oxford},\ \bibinfo {year} {2024})\ p.\ \bibinfo {pages}
  {298}\BibitemShut {NoStop}%
\end{thebibliography}
\end{document}